\documentclass[aps,prl,onecolumn,groupedaddress,showpacs]{revtex4}

\usepackage{graphicx}

\begin{document}

\draft 
\title{Landau-Zener problem for energies close to potential crossing points.}
 
\author{V.A.Benderskii} 
\affiliation {Institute of Problems of Chemical Physics, RAS \\ 142432 Moscow
Region, Chernogolovka, Russia} 
\affiliation{Laue-Langevin Institute, F-38042,
Grenoble, France} 
 
\author{E.V.Vetoshkin} 
\affiliation {Institute of Problems of Chemical Physics, RAS \\ 142432 Moscow
Region, Chernogolovka, Russia} 
\author{E. I. Kats} \affiliation{Laue-Langevin Institute, F-38042,
Grenoble, France} 
\affiliation{L. D. Landau Institute for Theoretical Physics, RAS, Moscow, Russia}
 
\date{\today}

\begin{abstract}
We examine one overlooked in previous investigations
aspect of well - known Landau - Zener (LZ) problem, namely, the behavior
in the intermediate, i.e. close to a crossing point, energy region, when all four LZ states
are coupled and should be taken into account. 
We calculate the $4 \times 4$ connection matrix in this intermediate energy region,
possessing the same block structure as the known
connection matrices for the tunneling and in the over-barrier regions of the energy,
and continously matching those in the corresponding energy regions.
Applications of the results may concern the various systems of 
physics, chemistry or biology, ranging from 
molecular magnets and glasses to Bose condensed atomic gases.

\end{abstract}

\maketitle

Standard textbook LZ theory \cite{LL65} treats of two linear diabatic potentials
$U^\# \pm F X$ crossing problem ($X=0$ is the crossing point).
However, in spite of more than half of century history, semiclassical
solutions of this problem have been found only in the limits of small or large energies $E$
(we will term these regions as tunneling or over-barrier, respectively), i.e. for
\begin{eqnarray} 
\label{jl1}
|U^\# - E| \gg U_{12} \, ; \, 
|U^\# - E| \ll U_{12} 
\, , 
\end{eqnarray} 
where $U_{12}$ is inter level interaction which in LZ model does not depend on $X$.
For the intermediate energy region
\begin{eqnarray} 
\label{jl2}
|U^\# - E| \leq U_{12} 
\,  
\end{eqnarray} 
there known only interpolating relations between exponentially decaying solutions in the tunneling 
and oscillating solutions in the over-barrier energy regions (see e.g. \cite{ZT01}, \cite{BV03}).
Analytical and numerical study of this region (\ref{jl2}) is the objective
of this paper. 

Our approach is motivated by semiclassical instanton approximation \cite{PO77}, \cite{CO85},
\cite{BM94}. The idea is to construct
two linearly independent continuous (with continuous first derivatives)
approximate solutions to the Schr\"odinger equation, which in the asymptotic
region coincide with semiclassic solutions, and in the vicinity of the
turning points - with the exact solutions of the so-called comparison
equation (i.e. the exact solution of the Schr\"odinger
equation for the chosen appropriately approximate near the turning
points potentials). In what follows the Weber equation \cite{EM53} 
will be used as the comparison equation, valid in the vicinity of the
second order turning points for an anharmonic potential \cite{BV03}, \cite{BV02}, \cite{BV04}.
To justify this choice it is sufficient to note that anharmonic corrections
remain semiclassically small (i.e., proportional
to higher orders of $\hbar $ series) in the region where the solutions of the comparison equation
have to be matched smoothly with the semiclassical solutions.
Luckily the analogous approach is valid to treat two diabatic potential crossing point
(LZ problem), and the comparison equations for this case are two coupled Weber equations with
the indices and arguments determined by the solutions of algebraic characteristic equation.

LZ problem for crossing diabatic potentials is equivalent to the coupled Schr\"{o}dinger equations 
which can be transformed by the substitution 
\begin{eqnarray} 
\label{nn11}
\Psi = \exp (\kappa X) \Phi  \,  
\end{eqnarray} 
into the 4-th order linear differential equation with independent of coordinates
coefficients at the derivatives
\begin{eqnarray} 
\label{apen1} 
D^4\Phi + 4\kappa D^3 \Phi + (6 \kappa ^2 - 2 \alpha \gamma ^2)D^2 \Phi
+ 4(\kappa ^3 - \alpha \gamma ^2 \kappa  - \frac{1}{2}\gamma ^2 f) D\Phi + 
\end{eqnarray}
\begin{eqnarray}
\nonumber
[\kappa ^4
- 2\alpha \gamma ^2 \kappa ^2 - 2 \gamma ^2 f \kappa + \gamma ^4 (\alpha ^2 - u_{12}^2 - f^2X^2)]\Phi = 0 
\, ,
\end{eqnarray}                                                                                                  
where $ D^n \equiv d^n/d X^n$, and $\gamma \gg 1$ 
is the dimensionless 
semiclassical parameter 
which is determined by the ratio of the characteristic potential scale over the
zero oscillation energy, and all other dimensionless appropriately rescaled variables are
\begin{eqnarray} 
\label{jl3}
\alpha = 2 \frac{U^\# - E}{\gamma \hbar \Omega } \, , \, f = \frac{2 a_0 F}{\gamma \hbar \Omega }
\, , \, u_{12} = \frac{2U_{12}}{\gamma \hbar \Omega }
\, , 
\end{eqnarray} 
where scale for the energy is given by $\Omega ^2 = F^2/m U_{12}$ ($m$ is a mass),  and space scale is determined
by the characteristic size $a_0$ of the potential in the vicinity of the crossing point.

The equation (\ref{apen1}) admits semiclassical solutions by 
Fedoryuk method \cite{FE64} -
\cite{FE66} since the coefficients at the $n$-th order derivatives proportional to 
$\gamma ^{-n}$, and
all the four asymptotic solutions read as  
\begin{eqnarray} 
\label{f1} 
\Psi _j^{(sc)} = (u_{12}^2 + f^2 X^2)^{-1/4} \exp \left (\int _{0}^{X} \lambda _j (x) d x \right )
\, , \, j = (++ , +- , -+ , --)
\, , 
\end{eqnarray} 
where we designated $\lambda _j = \lambda _j^0 + u_j $, $\lambda _j^0 = \pm \sqrt {\gamma (\alpha
\pm \sqrt {u_{12}^2 + f^2 X^2})}$, and $u_j = \gamma f ((\lambda _j^0)^2 - \alpha \gamma )^{-1}$.

The equation (\ref{apen1}), up to anharmonic terms proportional
to $X^2D \Phi $, $X^3 \Phi $, $X^4 \Phi $, can be formally derived by simple manipulations (two sequential differentiations
and summations) from the following second order equation
\begin{eqnarray} 
\label{apen2} 
D^2\Phi + (a_0 + a_1 X + a_2 X^2)\Phi = 0 
\, ,
\end{eqnarray}                                                                                                  
where the coefficients are
\begin{eqnarray} 
\label{apen3} 
a_0 = \kappa ^2 - \alpha \gamma ^2 - \frac{\gamma ^2 f}{2 \kappa }(1 + \delta )\, ;
\, a_1 = \gamma ^2f \delta \, ; \, a_2 = -\gamma^2 f\kappa \delta 
\, ,
\end{eqnarray}                                                                                                  
where the equation for $\kappa $ referred in what follows by the characteristic equation is  
\begin{eqnarray} 
\label{b8} 
\kappa ^4 - \alpha \gamma ^2\kappa ^2 + \frac{1}{4} \gamma ^4 u_{12}^2 = - \kappa ^4 \delta ^2(1 + 2\delta )
+ R(\kappa , \delta )\, ,
\end{eqnarray} 
where
\begin{eqnarray}
\nonumber
R(\kappa , \delta ) = (2 \kappa ^6)^{-1}(1-3\delta )(1+\delta )^{-3}(1 - Q - \sqrt {1 - 2Q^2)})\, ; \, 
Q=8\delta ^2(1+\delta )\, ,
\end{eqnarray}
and
\begin{eqnarray} 
\label{b9} 
\delta = \frac{\gamma ^2 f}{4\kappa ^3 } \, , 
\end{eqnarray} 
The fundamental solutions to (\ref{apen2}) read as
\begin{eqnarray} 
\label{apen4} 
D_p\left [\pm \left (\frac{\gamma ^4 f^2}{\kappa ^2}\right )^{1/4} \left (X - \frac{1}{2\kappa }\right )\right ]
\, ,
\end{eqnarray}                                                                                                  
where
\begin{eqnarray} 
\label{apen5} 
p = -\frac{1}{2} + \left (\frac{\gamma ^4 f^2}{\kappa ^2}\right )^{-1/2} \left (a_0  - \frac{a_1^2}{4 a_2}\right ) 
\, .
\end{eqnarray}                                                                                                  
In the tunneling and over-barrier regions of energies, where $\delta < 1/4$, 
these 4 solutions (2 solutions of (\ref{apen4}) for two largest modulus roots of
the characteristic equation (\ref{b8})) can be separated into two independent pairs
(orthogonality of the Weber functions with different indices). Thus one can say that
the crossing point is equivalent to two second order turning points with different Stokes
constants (see e.g., \cite{HE62}).

In the tunneling region the two largest modulus roots of (\ref{b8}) are
(two other roots are small and do not satisfy semiclassical approach)
\begin{eqnarray} 
\label{apen6} 
\kappa = \pm \kappa _0\left (1 \pm \frac{\delta ^2}{2}\frac{\kappa _0^2}{2\kappa _0^2 - \alpha \gamma ^2}\right )
\, ; \,
\kappa _0 = \frac{\gamma }{\sqrt 2}\left (\alpha + \sqrt {\alpha ^2 - u_{12}^2}\right )^{1/2} 
\, ,
\end{eqnarray}
and the four linearly independent solutions (\ref{apen4}) are matched to the semiclassical
solutions 
(\ref{jl3}) in the region $\alpha > f |X| > u_{12}$, where the exponent of the wave
function can be expanded over small parameter $\delta $
\begin{eqnarray} 
\label{jl4} 
\Phi \propto \exp \left (\kappa X + \delta (\kappa X)^2 - \frac{2}{3}\delta ^2 (\kappa X)^3 + .....\right )
\, .
\end{eqnarray}
Since $\kappa |X| \leq (4\delta )^{-1}$, the convergent with alternating signs expansion (\ref{apen6})
determines the accuracy of the asymptotically smooth transformation.
Putting altogether we end up at the conclusion
that anharmonic corrections
to the Weber functions (\ref{apen4}) are small (by other words the parameter $\delta $ determines the accuracy 
of our approximation).
The same kind of analysis can be performed in the over-barrier region, where one finds
two imaginary largest modulus roots of the characteristic equation (see details in \cite{BV04}).

More difficult task is to find solutions in the intermediate
energy region, where two roots of the characteristic equation 
are real and two are imaginary ones having the same modulus, i.e. moving upon $\alpha $ variation along a circle
with the radius $\gamma \sqrt {u_{12}/2}$. In this case the semiclassical solutions can be presented as certain
linear combinations of the comparison equation solutions, and
the roots are
\begin{eqnarray} 
\label{apen9} 
\kappa _{1 , 2} \simeq \pm \gamma \sqrt {\frac{u_{12}}{2}}\exp (i\varphi ) 
\, ; \,
\kappa _{3 , 4} \simeq \pm i \gamma \sqrt {\frac{u_{12}}{2}}\exp (-i\varphi )
\, , 
\end{eqnarray}                                                                                                  
where
\begin{eqnarray} 
\label{apen10} 
\tan \varphi = \sqrt {\frac{u_{12}- \alpha }{u_{12} + \alpha }}
\, . 
\end{eqnarray}                                                                                                  
Correspondingly to these roots (\ref{apen10}) the arguments and the indices
of the Weber functions (\ref{apen4}), (\ref{apen5}) read as
\begin{eqnarray} 
\label{apen11} 
z_1 = z_2 = 2 \kappa _{int}\sqrt {\delta _{int}} \exp (- i \varphi /2)(X - (2 \kappa _{int})^{-1}\exp (-i \varphi )
\, ; \,
\end{eqnarray}
\begin{eqnarray}
\nonumber
z_3 = z_4 = 2 \kappa _{int}\sqrt {\delta _{int}} \exp (i \varphi /2)(X - (2 \kappa _{int})^{-1}\exp (i \varphi )
\, , 
\end{eqnarray}                                                                                                  
and
\begin{eqnarray} 
\label{apen12} 
p_1 = p_2 - 1 = -1 - \frac{1}{4 \delta _{int}} \exp (- i \varphi )(1 + 2\delta _{int}^2 \exp (-2 i \varphi ))
\, ;\,
\end{eqnarray}
\begin{eqnarray}
\nonumber 
p_4 = p_3 - 1 = -1 - \frac{1}{4 \delta _{int}} \exp (i \varphi )(1 + 2\delta _{int}^2 \exp (2 i \varphi ))
\, , 
\end{eqnarray}                                                                                                  
where 
\begin{eqnarray}
\label{new1}
\kappa _{int} = \gamma (u_{12}/2)^{1/2} \, ; \,
\delta _{int} = (\gamma ^2 f)/(4 \kappa _{int}^3)
\, .
\end{eqnarray}
The semiclassical solutions (\ref{f1}) are matched asymptotically 
smoothly  to the linear combinations
of the Weber functions in the region $u_{12} > f |X|$. Since from (\ref{apen11}), (\ref{apen12})
follow that at large $\delta _{int}$ the indices of the Weber functions are also large,
one can use known due to Olver \cite{OL59}, \cite{OL74}
asymptotics of the Weber functions 
with
large arguments and indices 
\begin{eqnarray} 
\label{apen15} 
D_p(z) \propto \exp \left [ - \frac{1}{2} \int \left (z^2 - 4\left (p + \frac{1}{2}\right )\right )^{1/2} dz \right ]
\, .
\end{eqnarray}
At $z^2 \gg 4|p + (1/2)|$ (\ref{apen15}) is reduced to the usual asymptotic expansion
of the large argument Weber functions, and in the opposite limit (i.e. in the intermediate region)
(\ref{apen15}) corresponds to the expansion of the exponent over odd powers of $z$.
We can also find asymptotics to the solutions of (\ref{apen2})
\begin{eqnarray} 
\label{apen16} 
\Phi _0 \propto \exp \left (-i \int \sqrt {a_0 + a_1X + a_2X^2} dx \right )
\, 
\end{eqnarray}                  
valid at arbitrary values of the parameters $a_i$ ($a_2=0$ including).

The fundamental solutions to the comparison equation are the asymptotics for the wave functions
in the form (\ref{nn11}), namely,
\begin{eqnarray} 
\label{jl11}
\Psi ^\pm _j  = \exp (\kappa X) D_p (\pm z_j(X)) \, ,  
\end{eqnarray} 
and
\begin{eqnarray} 
\label{jl12}
\Psi ^+_1 , \Psi ^+_4 \propto \exp(F_1(X)) \, , \,
\Psi ^-_2 , \Psi ^-_3 \propto \exp(-F_1(X)) \, , \,
\Psi ^-_1 , \Psi ^-_3 \propto \exp(iF_2(X)) \, , \,
\Psi ^+_2 , \Psi ^-_4 \propto \exp(-iF_2(X)) \, , \,
\, ,
\end{eqnarray} 
where
\begin{eqnarray} 
\label{jl13} 
F_{1 , 2}(X) = \gamma \sqrt {u_{12} \pm \alpha }(1 + \delta _{int}) X  
- \kappa _{int}^2 \delta _{int}^2 \exp(-2i \varphi ) X^2 +
\frac{\gamma f^2}{12 u_{12} \sqrt {u_{12} \pm \alpha }}
\left (1 \pm \frac{\alpha }{u_{12}} - \delta _{int}\right ) X^3
\, .
\end{eqnarray}                                                                                                  
The wave functions (\ref{jl12}) asymptotically smoothly turn into the
semiclassical functions (\ref{f1}). The accuracy of this matching is determined
by the anharmonic corrections, i.e. by the parameter $\delta _{int}$ (\ref{new1}),
and the Olver asymptotic (\ref{apen15}) works even on the boundary  
$z^2 \simeq 4|p + (1/2)|$. The parameter $\delta _{int}$ is no more small one,
when simultaneously
\begin{eqnarray} 
\label{jl14} 
|\alpha | \leq \left (\frac{f}{\gamma }\right )^{2/3} =\left (\frac{u_{12}}{2 \gamma ^2}\right )^{1/3}
\, ; \,
u_{12} \leq \frac{2}{\gamma } 
\, .
\end{eqnarray}                                                                                                  
However the asymptotic matching of the solutions should be performed at small $|X| < \gamma ^{-1}$,
where the comparison equation (\ref{apen5}), and, therefore, the characteristic equation (\ref{b8})
are valid, even though upon increasing $\delta _{int}$ the potential becomes more and more anharmonic one.
At $\alpha =0$, and $u_{12}=0$, the equation (\ref{b8}) (with the term $R(\kappa , \delta)$
taking into account)
has the doubly degenerate root $\kappa =0$,
i.e. in terms of (\ref{apen5}) $a_1 = \pm \gamma ^2 f$, $a_2=0$. Thus in this limit (\ref{apen5})
is equivalent to two decoupled Airy equations, corresponding to the diabatic potentials. These
solutions turns smoothly into the semiclassical ones (\ref{f1}), and the anharmonic corrections
in the matching region are small over the parameter $\gamma ^{-1/2}$.

We conclude that in the both intermediate energy subregions: large $\kappa $ (i.e. $\propto \gamma $), and small
small $\kappa $ (i.e. $\propto \sqrt \gamma $) the comparison equation (\ref{apen5}) is reduced
to two decoupled equations, Weber or Airy ones, respectively.
This simple observation enables us to construct the universal connection matrix for the
both intermediate energy region by using Olver asymptotic expansion 
(\ref{apen16}).
The four roots (\ref{apen9}) distributed over the circle with the radius $\gamma \sqrt {(u_{12}/2)}$
on the comples plane determine the following combinations of the comparison
equation solutions matching the semiclassical solutions (\ref{f1}).
Namely
\begin{eqnarray} 
\label{jl15} 
\Psi _1^+ + \Psi _4^+ \to \Psi _{++}^{sc} \, ; 
\, \Psi _2^- + \Psi _3^- \to \Psi _{+-}^{sc} \, ; 
\, \Psi _1^- + \Psi _3^+ \to \Psi _{-+}^{sc} \, ; 
\, \Psi _2^+ + \Psi _4^- \to \Psi _{--}^{sc} 
\, ,
\end{eqnarray}                                                                                                  
where the superscript $sc$ points out the semiclassical solutions.
Combining together the asymptotic expansions for these
combinations, we find 
at the crossing point, the matrix 
$\hat U_c^{\prime \prime }$ is 
\begin{eqnarray} && 
\label{q888} 
\hat {U}_c^{\prime \prime } = \left [ 
\begin{array}{cc} 
(\sqrt {2\pi }/\Gamma (q^*))\exp (-2\chi (q^*)) & 0   \\
0 & (\Gamma (q)/\sqrt {2\pi })\exp (2\chi (q))(1 - \exp (-2\pi q_2)\cos ^2(\pi q_1)) \\
0 & \exp (-2\pi q_2)\cos (\pi q_1)   \\
\exp(-2\pi q_2)\cos (\pi q_1) & 0  
\end{array} 
\right .  
\end{eqnarray}
\begin{eqnarray}
\nonumber
\left .
\begin{array}{cc}
0 & -\exp (-2\pi q_2)\cos (\pi q_1)  \\
-\exp (-2\pi q_2)\cos (\pi q_1)  &  
0  \\
(\sqrt {2\pi }/\Gamma (q))\exp (2\chi (q)) & 0 \\
0 & (\Gamma (q^*)/\sqrt {2 \pi })\exp (2\chi (q^*))(1 - \exp (-2\pi
q_2)\cos ^2(\pi q_1)) 
\end{array} 
\right ] \, , 
\end{eqnarray} 
where 
\begin{eqnarray}
\label{a55}
q  = q_1 + i q_2\, ; \, 
q_{1 , 2} = \frac{\gamma u_{12}\sqrt {u_{12} \pm \alpha }}{4 f} \,
; \, q^* = q_1 - i q_2 \, ,
\end{eqnarray}
and, besides, we introduce the following abridged notations
\begin{eqnarray}
\label{a551}
\chi = \chi _1 + i \chi _2 \, ; \, 2 \chi _1 = q_1 - \left (q_1 - \frac{1}{2}\right ) \ln |q| + \varphi q_2 \, 
,
\end{eqnarray}
and analogously 
\begin{eqnarray}
\label{a552} 
2\chi _2 = q_2 - q_2\ln |q| - \varphi \left (q_1 - \frac{1}{2}\right ) 
\, , 
\end{eqnarray}
where $\varphi $ is defined by 
(\ref{apen10}). 

The connection matrix (\ref{q888}) in the intermediate energy region
is our main result and the motivation of this publication.
This matrix generalizes the results we presented in \cite{BV03}.
It is ready for further applications and to reap the fruits of the result we compute
the LZ transition probability $|T|^2$ universally valid for the tunneling, over-barrier and intermediate energy
regions (solid line in the Fig. 1). 
It is instructive to compare our result with the perturbative Landau
approach (see e.g., \cite{LL65}) valid at small coupling constants.
In the first order perturbation theory the transition amplitude $A_{LZ}^{(1)}$
reads
\begin{eqnarray}
\label{jl552} 
A_{LZ}^{(1)} = 2 i u_{12} \sqrt {\frac{\pi }{\varepsilon }} Ai(2^{2/3} \varepsilon ) 
\, , 
\end{eqnarray}
where we designated $\varepsilon = -\alpha f^{-2/3}$, and $Ai$ is the first kind
Airy function.
All even higher order terms equal zero, and odd terms read as
\begin{eqnarray}
\label{jl553} 
A_{LZ}^{(2n + 1)} = (A_{LZ}^{(1)})^{2n +1} 
\, .
\end{eqnarray}
Therefore the series can be easily sum up giving the generalized Landau formula
\begin{eqnarray}
\label{jl554} 
A_{LZ} = A_{LZ}^{(1)}\left [ 1 + (A_{LZ})^2\right ]^{-1} 
\, .
\end{eqnarray}
We show this perturbative solution by the dashed line on the Fig. 1.
Note that although 
the equation (\ref{jl553}) reproduces the oscillating energy dependence of the LZ transition
amplitudes, the equation gives the period of the oscillations
which are quite different from those we calculated by
our $4 \times 4$ connection matrix (\ref{q888}).
The difference occurs because the perturbation (say $2 \time 2$) method disregards the contributions
from the increasing solutions to the Schr\"odinger equation, which are relevant
in the intermediate energy region.

\acknowledgements 
The research described in this publication was made possible in part by RFFR Grants. 
One of us (E.K.) is indebted to INTAS Grant (under No. 01-0105) for partial support,
and V.B. and E.V. are thankful to CRDF Grant RU-C1-2575-MO-04.

\newpage

\centerline{Figure Captions.}

Fig. 1

Energy dependent LZ transition probability:

Solid line - the $4 \times 4$ connection matrix (\ref{q888}) calculations;  

Dashed line - the generalized perturbative Landau formula (\ref{jl554}).

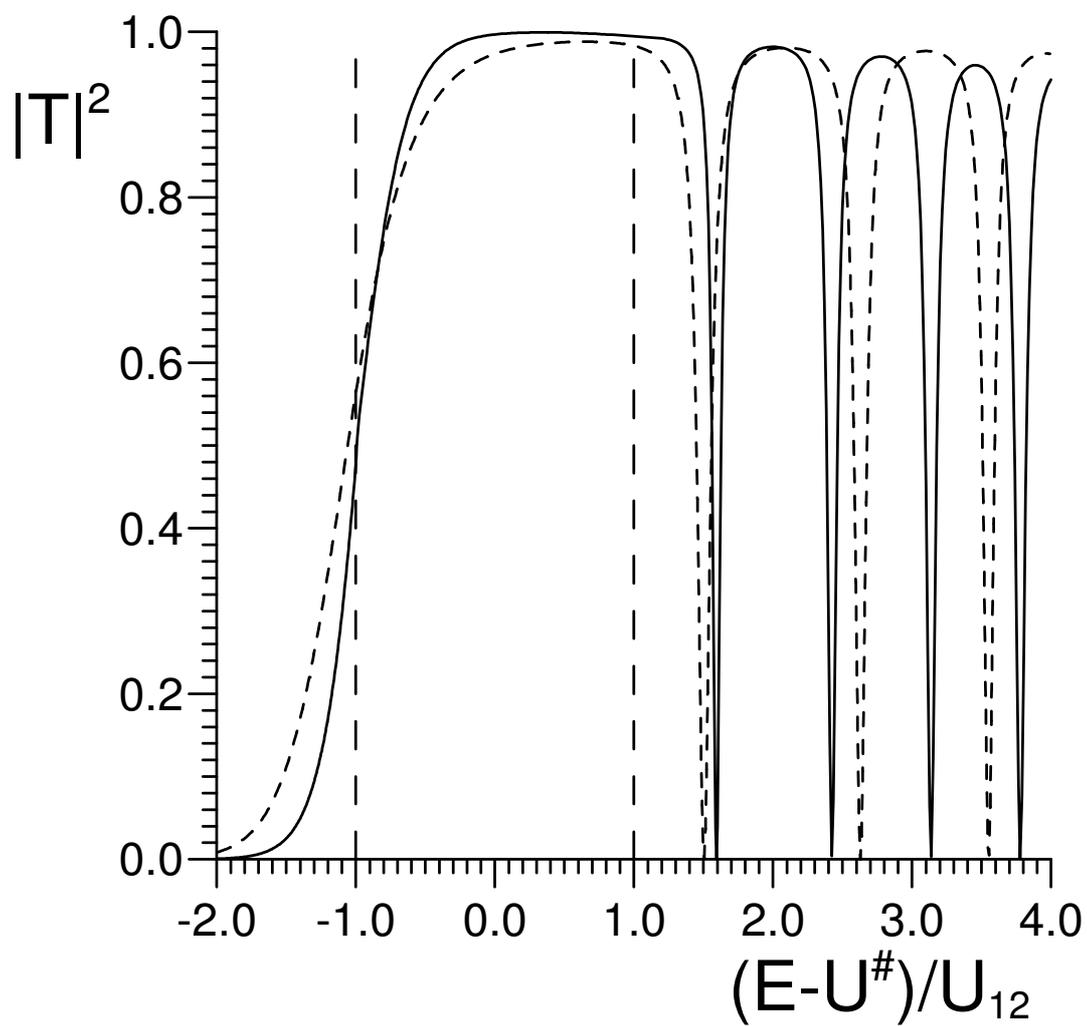
\end{document}